\documentclass{elsart}
\newcommand{\pdf}{n}
\usepackage{citesort}
\usepackage{amssymb}
\usepackage{ifthen}
\ifthenelse{\equal{\pdf}{y}}
{\pdfoutput=1 \pdfcompresslevel=9
\usepackage[pdftex]{graphicx}}
{
\usepackage{graphicx}
}
\begin{document}
\DeclareGraphicsExtensions{jpg,gif,tif,png}
\begin{frontmatter}
\title{Long-time behavior of Granular Gases\\ with impact-velocity dependent\\ coefficient of restitution}
\author{Thorsten P\"oschel, Nikolai V. Brilliantov and Thomas Schwager}
\address{Humboldt-Universit\"at zu Berlin -- Charit\'e, Institut f\"ur Biochemie,\\
  Monbijoustra{\ss}e 2, D-10117 Berlin, Germany}
\begin{abstract}
A gas of particles which collide inelastically if their impact velocity exceeds a certain value is investigated. In difference to common granular gases, cluster formation occurs only as a transient phenomenon. We calculate the decay of temperature due to inelastic collisions. In spite of the drastically reduced dissipation at low temperature the temperature surprisingly converges to zero. 
\end{abstract}

\end{frontmatter}

\section{Introduction}
\label{sec:intro}
Gases of dissipatively colliding particles (Granular Gases) in the absence of external forces reveal a variety of interesting phenomena, such as characteristic deviations from the Maxwell distribution \cite{GoldshteinShapiro1:1995,NoijeErnst:1998,BrilliantovPoeschelStability:2000}, overpopulation of the high energy tail of the distribution function \cite{EsipovPoeschel:1995}, anomalous diffusion \cite{BrilliantovPoeschel:1998d,BreyRuizMonteroCuberoGarcia:2000}, and others (see \cite{PoeschelLuding:2001} for an overview). The most striking phenomenon which makes Granular Gases different from molecular gases is the self-organized formation of spatio-temporal structures such as clusters \cite{GoldhirschZanetti:1993} and vortices \cite{BritoErnst:1998}. 

The loss of mechanical energy of dissipatively colliding particles $i$ and $j$ is characterized by the (normal) coefficient of restitution, which relates the normal component of the relative velocity before a collision, $g$, to that after the collision, $g^\prime$:
\begin{equation}
  \label{eq:epsdef}
  \varepsilon\equiv \frac{g^\prime}{g}=-\frac{\vec{v}_{ij}^{\,\prime}\cdot \vec{e}_{ij}}{\vec{v}_{ij} \cdot \vec{e}_{ij}}\,,~~~~~~~~~~~~~~~~~~~~~~~~~~
\vec{e}_{ij}\equiv \frac{\vec{r}_i-\vec{r}_j}{\left|\vec{r}_i-\vec{r}_j\right|}
\end{equation}
with $\vec{v}_{ij}\equiv \vec{v}_i-\vec{v}_j$ and $\vec{v}_{ij}^{\,\prime}$ being the corresponding after-collision value. Frequently it is assumed that the coefficient of restitution is a material constant which simplifies analytical calculations considerably. This assumption, however, contradicts experiments \cite{BridgesHatzesLin:1984} and disagrees with a dimension analysis \cite{Tanaka,Ramirez:1999}. Instead the coefficient of restitution is a function of the impact velocity.

Obviously, the coefficient of restitution can be obtained by integrating Newton's equation of motion for the collision with appropriate choice of the contact force. The elastic component of the contact force of spheres of diameter $\sigma$ is given by Hertz' law $F^{({\rm el})}=B\xi^{3/2}$, $\xi(t)\equiv\sigma-\left|\vec{r}_i-\vec{r}_j\right|$ with $B(\sigma)$ being the elastic material parameter \cite{Hertz:1882}. Assuming viscoelastic material properties the dissipative part of the contact force reads $F^{({\rm dis})}=A\sqrt{\xi}\dot{\xi}$  \cite{BrilliantovSpahnHertzschPoeschel:1994,MorgadoOppenheim:1997,KuwabaraKono:1987} with the dissipative material constant $A(\sigma)$. Integration of Newton's equation of motion yields then \cite{SchwagerPoeschel:1998,Ramirez:1999}
\begin{equation}
  \label{eq:epsvisco}
  \varepsilon_v(g)= 1-D_1 g^{1/5}+ D_2 g^{2/5}\mp\dots\,,
\end{equation}
where the coefficients $D_{1/2}$ depend on the elastic and dissipative particle material properties and on the radii of the particles (for details see \cite{SchwagerPoeschel:1998}). 

The temperature of Granular Gases in the absence of spatial correlations is defined as the mean kinetic energy of the particles, just as for molecular gases. Due to dissipative collisions the temperature decays persistently and, therefore, the thermal velocity $v_T(t)\equiv \sqrt{2T(t)/m}$ decays too. For vanishing temperature the particles collide elastically, i.e., $\lim_{T\to 0}\varepsilon=1$. Since the formation of clusters in Granular Gases is a consequence of the dissipative particle collisions the question arises whether for gases of viscoelastic particles clusters may occur and persist. The linear stability analysis of the hydrodynamic equations of such gases supports the speculation that the cluster state is only a transient phenomenon for the case of viscoelastic particles  \cite{BrilliantovPoeschel:2001roy}.

For principal reasons (to be discussed below) the numerical investigation of the long time behavior of force free Granular Gases of viscoelastic particles is problematic. Therefore, in this paper we assume a coefficient of restitution with an extremely simplified impact velocity dependence: The particles collide with a constant coefficient of restitution $\varepsilon^*$ if the impact velocity exceeds a certain value $g^*$, otherwise they collide elastically:
\begin{equation}
  \label{eq:epsstep}
  \varepsilon(g)=\left\{
    \begin{tabular}{ll}
      $\varepsilon^*$ & for $g>g^*$\\
      1 & for  $g \le g^*$\,.
    \end{tabular}
\right.
\end{equation}
We wish to mention that the coefficient of restitution of the form (\ref{eq:epsstep}) was used in \cite{NieBenNaimChen2002} as a numerical trick to avoid the inelastic collapse in event driven MD simulations. In this context the transition velocity was chosen to be much smaller than the thermal velocity, $g*\ll v_T$. 
Before investigating the cluster formation process first we will derive the law of temperature decay in the homogeneous cooling state.

\section{Decay of temperature in the homogeneous cooling state}

In the absence of external forces a Granular Gas which is initialized at uniform number density $n$ stays homogeneous during the first stage of its evolution, called {\em homogeneous cooling state}. In this regime the kinetic energy of the gas may be characterized by the granular temperature $T$, defined by the second moment of the velocity distribution function $f(v)$
\begin{equation}
  \label{eq:Tdef}
  \frac{d}{2}n T \equiv  \int \frac{mv^2}{2}f(v) d \vec{v} \, , 
\end{equation}
($m$ is the mass of particles and $d$ is the dimension) which decays persistently due to inelastic collisions. For $\varepsilon=\mbox{const.}$ the decay is described by Haff's law $T_H(t)$ \cite{Haff:83}, whereas for gases of viscoelastic particles the temperature is given by $T_v(t)$  \cite{SchwagerPoeschel:1998}
\begin{equation}
  \label{eq:Haff}
  T_H(t)=\frac{T_0}{\left(1+t/\tau_H\right)^2},\qquad\qquad  T_v(t)=\frac{T_0}{\left(1+t/\tau_v\right)^{5/3}}\,,
\end{equation}
where $\tau_H$ and $\tau_v$ are the relaxation times \cite{BrilliantovPoeschel:2000visc}. 

Equations (\ref{eq:Haff}) have been first derived with the assumption of a Maxwell distribution for the particle velocities. It is known that the velocity distribution function of Granular Gases deviates slightly from a Maxwell distribution \cite{GoldshteinShapiro1:1995,NoijeErnst:1998,BrilliantovPoeschelStability:2000}, however, the functional form of Eqs. (\ref{eq:Haff}) keeps conserved also when we take into account these deviations. Before investigating the temperature decay let us first check whether the assumption of an approximative Maxwell distribution is justified for our case too. 

Starting at a certain temperature which corresponds to a thermal velocity $v_T\gg g^*$ almost all collisions occur with a constant coefficient of restitution $\varepsilon^*$, hence, in this range we expect to find in the homogeneous cooling state a velocity distribution close to a Maxwell distribution with the small deviations mentioned above. At late times when the thermal velocity is small with respect to the transition velocity, $v_T\ll g^*$, the majority of the collision occurs elastically. In this range we expect, therefore, that the particle velocities obey a Maxwell distribution just as for molecular gases. In the intermediate range $v_T\approx g^*$ a sizable part of the collisions, i.e. collisions between slow particles, occurs elastically, whereas fast particles mainly undergo dissipative collisions. Therefore, it is not \`a priori clear whether the distribution function is close to the Maxwell distribution.

\begin{figure}[htbp]
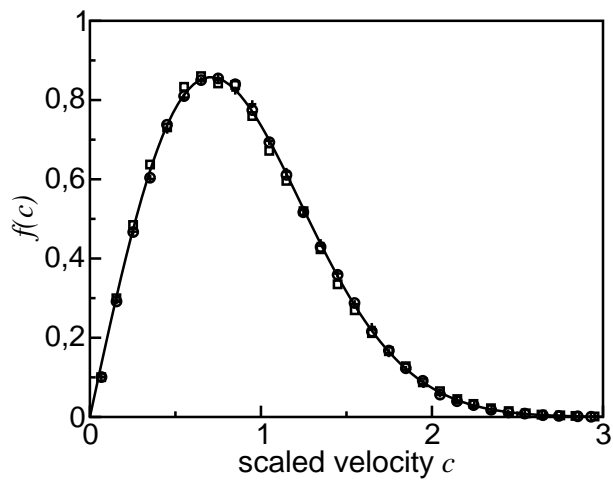

\ifthenelse{\equal{\pdf}{y}}
{  \centerline{\includegraphics[width=8cm]{figs/distr.jpg}} }
{  \centerline{\includegraphics[width=8cm,clip=]{figs/distr.eps}} }
  \caption{Scaled velocity distribution function obtained by MD simulations at times $t_1<t_2<t_3$ where $v_T(t_1)\gg g^*$ (circles), $v_T(t_2)\approx g^*$ (squares), and $v_T(t_3)\ll g^*$ (plus symbols). The full line shows a Maxwell distribution.}
  \label{fig:distr}
\end{figure}

We have simulated a two-dimensional force free cooling Granular Gas of $N=10^5$ particles of unit mass which collide according to the rule (\ref{eq:epsstep}) by event-driven Molecular Dynamics. The parameters are: $\varepsilon^*=0.6$, $g^*=0.1$, $\eta=0.1$,  and the initial temperature is $T(0)=1$ which corresponds to an initial thermal velocity $v_T(0)=\sqrt{2}$. Figure \ref{fig:distr} shows the scaled velocity distribution function $\tilde{f}(c)$
\begin{equation}
f(\vec{v},t)=\frac{n}{v_T^d(t)} \tilde{f}\left(\vec{c}\,\right)\, ,
\qquad\qquad \vec{c} \equiv \frac{\vec{v}}{v_T(t)} \, , 
\end{equation}
for the three mentioned cases together with the Maxwell distribution. From the results we conclude that even in the transition region $v_T\approx g^*$ the Maxwell distribution is a good approximation for the velocity distribution function. 

Using the standard approach based on the Boltzmann equation (e.g. \cite{GoldshteinShapiro1:1995,NoijeErnst:1998,BrilliantovPoeschelStability:2000,BrilliantovPoeschel:2000visc}) with the assumption of a Maxwell distribution the equation for the time evolution of temperature may be obtained
\begin{equation}
\label{eq:dTdt}
\frac{dT}{dt}= -\frac{2}{d}g_2(\sigma) \sigma^{d-1} n v_T T \mu_2 \, , 
\end{equation}
where $\sigma$ is the diameter of the particles,  $g_2(\sigma)$ is the contact value of the pair distribution function ($g_2(\sigma) = (1-7 \eta/16)/(1-\eta)^2$ for a two-dimensional gas with packing fraction $\eta= \pi \sigma^2 n /4$), and $\mu_2$ is the second moment of the dimensionless collision integral, which may be written in the form  \cite{BrilliantovPoeschelStability:2000,BrilliantovPoeschel:2000visc}
\begin{equation}
\label{eq:def_mu2}
\mu_2= \frac14 \int d\vec{c}_1 \int d\vec{c}_2 \int d\vec{e}\, \Theta(-\vec{c}_{12} \cdot \vec{e}\, ) 
\tilde{f}(\vec{c}_1) \tilde{f}(\vec{c}_2) (1-\varepsilon^2) \left|  \vec{c}_{12} \cdot \vec{e}\, \right|^3 \, .
\end{equation}
The integration is performed over the velocities $\vec{c}_1$, $\vec{c}_2$ of the colliding pair and over the unit vector $\vec{e}=\vec{e}_{12}$ as introduced in Eq. (\ref{eq:epsdef}). The unit step function $\Theta(x)$ guarantees that collisions occur only between approaching particles. 

Approximating the velocity distribution function by the Maxwell distribution, $\tilde{f}(\vec{c}\,)=\pi^{-d/2} \exp(-c^2)$, the above moment may be calculated. For two-dimensional systems it reads
\begin{equation}
\label{eq:mu2_fin}
\mu_2= \frac{\sqrt{2 \pi}}{2} \left( 1-\varepsilon^{* \, 2} \right) \left(1+ \frac{g_0^2}{2u} \right) 
\exp \left( - \frac{g_0^2}{2u} \right) \, , 
\end{equation}
where $u=u(t)=T(t)/T_0$ is the reduced temperature ($T_0=T(0)$) and $g_0=g^*/v_T(0)$. 
Introducing the variable $x=2u/g_0^2$ we obtain from Eq. (\ref{eq:dTdt})
\begin{equation}
\label{eq:xdot}
\dot{x}= -\alpha \left( x^{3/2} + x^{1/2}  \right) e^{-1/x} 
\end{equation}
where 
\begin{equation}
\label{eq:def_alpha}
\alpha= g_0 \left( 1-\varepsilon^{* \, 2} \right)  g_2(\sigma) \sigma n \sqrt{\frac{\pi T_0}{2 m}}\, . 
\end{equation}
If $v_T(0) \gg g^*$, i.e. $g_0 \ll 1$, then $x \gg 1$ at the initial stage of the temperature evolution. Equation (\ref{eq:xdot}) reduces then to $\dot{x}=-\alpha x^{3/2}$, yielding Haff's law (\ref{eq:Haff}) with 
\begin{equation}
  \label{eq:tauH}
        \tau_H = \frac{\alpha}{g_0 \sqrt{2}} = \frac12 \left( 1-\varepsilon^{* \, 2} \right) g_2(\sigma) \sigma n \sqrt{\frac{\pi T_0}{m}} \, .
\end{equation}
For the late stage of the temperature evolution, i.e. for sufficiently small $x$, the general solution of Eq. (\ref{eq:xdot}) can be found as a series
\begin{equation}
\label{eq:solut_x}
\frac{x^{3/2}e^{1/x}}{(1+x)}
\left[ 1+ \frac12 \frac{(3+x)}{(1+x)}x +\frac14 \frac{(15+10x+3x^2)}{(1+x)^2} x^2 +\dots \right] 
= \alpha t + C \, ,  
\end{equation}
where $C$ is an integration constant. For large times, i.e., $x\ll1$ it simplifies to
\begin{equation}
\label{eq:small_x}
x^{3/2}e^{1/x} = \alpha t
\end{equation}
Applying the (implicit) definition of Lambert's $W$ function \cite{Corless:1996}
\begin{equation}
  W(x)e^{W(x)}=x
\end{equation}
we find the analytical solution of Eq. (\ref{eq:small_x})
\begin{equation}
  \frac{1}{x} = -\frac{3}{2}W\left(-\frac{2}{3(\alpha t)^{2/3}}\right)~;\qquad x=\frac{2}{g_0^2}\frac{T}{T_0}
  \label{eq:asymptotics}
\end{equation}
For large values of $t$ we get the approximation (choosing the appropriate branch of $W(x)$)
\begin{equation}
  \frac{1}{x}= \frac{3}{2}\left[\ln\left(\frac{3}{2}(\alpha t)^{2/3}\right) + \ln\ln\left(\frac{3}{2}(\alpha t)^{2/3}\right)+\dots\right]
\label{eq:xasympt}
\end{equation}
which yields the asymptotics for $t\to\infty$
\begin{equation}
\label{eq:Temp}
T(t) = \frac{g_0^2}{2} \frac{T_0}{\log \alpha t}\, ,    
\end{equation}
with the constants $g_0$ and $\alpha$ defined above. 

\begin{figure}[htbp]
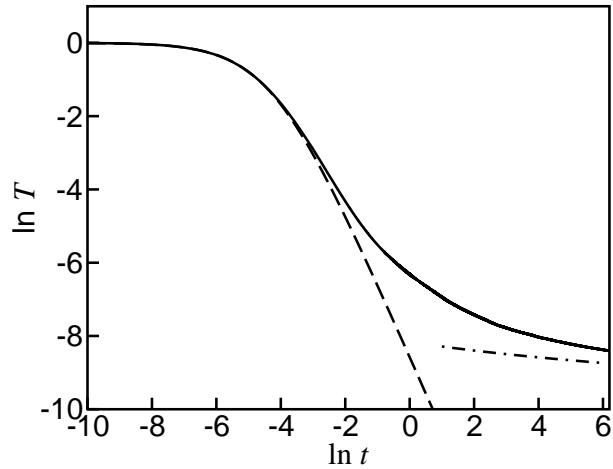

\ifthenelse{\equal{\pdf}{y}}
{    \centerline{\includegraphics[width=8cm]{figs/Temp.jpg}}}
{    \centerline{\includegraphics[width=8cm,clip=]{figs/Temp.eps}}}
  \caption{Evolution of the granular temperature. Full line: MD simulation, dashed line: Haff's law (\ref{eq:Haff}), dash-dotted line: asymptotics due to Eq. (\ref{eq:asymptotics}).}
  \label{fig:Temp}
\end{figure}

At late times most of the particles collide elastically due to Eq. (\ref{eq:epsstep}), therefore, only collisions of particles whose velocities belong to the high-energy tail contribute to the decay of temperature. For this reason the asymptotic temperature relaxation is extremely slow as compared with the power laws Eq. (\ref{eq:Haff}), see Fig. \ref{fig:Temp}. 

For Granular Gases of particles which collide with $\varepsilon=\mbox{const.}$ it is known that the high energy tail of the scaled distribution function does not obey a Maxwell distribution ($\sim\!\!\exp(-c^2)$) but decays as $\exp(-{\rm const.} \times c)$ \cite{EsipovPoeschel:1995}. This overpopulation of the high energy tail arises due to a mechanism which is specific for gases of dissipative particles: 
the scaled velocity of a particle, $c=v/v_T$  can increase either by collisions with other particles or by keeping its velocity $v$ while the thermal velocity $v_T$, i.e. temperature, decreases. Both processes cause increase of the number of fast particles, i.e., they enhance the population of the high velocity tail. For very fast particles, $c \gg 1$,  the former process may be neglected and the latter one leads to the exponential distribution function $\exp(-{\rm const.} \times  c)$ \cite{EsipovPoeschel:1995}. We wish to stress that for $\varepsilon=\mbox{const.}$ the Granular Gas has no inherent time scale and the shape of the scaled velocity distribution function $\tilde{f}(c)$ does not depend on time. Hence, the exponential overpopulation of the high-velocity tail is a time-independent effect. 

For the coefficient of restitution according to Eq. (\ref{eq:epsstep}) the overpopulation of the tail will not be observed at late times. The temperature decays only logarithmically slow as compared with the power law for $\varepsilon=\mbox{const}$. Therefore, the process of depopulation of the tail due to collisions is not balanced anymore by the decrease of the temperature of the gas, which enhances the tail population.  According to this argument in difference to the case $\varepsilon=\mbox{const.}$ we expect an {\em underpopulation} of the high-energy tail.
There is a non-negligible probability that particles collide in such a way that the after-collision velocity of one of the collision partners exceed $g^*$ even if the pre-collision velocities of both partners are smaller than $g^*$. The probability of such collisions, which are responsible for the population of the velocity range  $v^\prime > g^*$, decays rapidly as a function of the after-collision velocity $v^{\prime}$. To estimate the affected range of particle velocities we consider the collision of {\em elastic} particles: The maximal after-collision velocity which may emerge in a collision of elastic particles with pre-collision velocity $g^*$ is $\sqrt{2}g^*$. Hence we expect that only the interval of velocities $(g^*, \sqrt{2}g^*)$ will be significantly populated. 

At present it remains unclear whether the deviation from the Maxwell distribution at high velocities does change the behavior of the gas at late times significantly.  This intriguing problem is subject of current research. 
%\vfill\newpage

\section{Cluster formation}   

The spontaneous formation of clusters in a force free cooling Granular Gas can be understood by simple arguments \cite{GoldhirschZanetti:1993}: Consider fluctuations of the density in an otherwise homogeneous Granular Gas. In denser regions the particles collide more frequently than in more dilute regions, therefore, dense regions of the system cool faster than dilute regions and the local pressure decays in these colder regions. The resulting pressure gradient causes a flux of particles into the denser region, which leads to further increase of the density. Hence, small fluctuations of the density are enhanced which leads to the formation of clusters. 

These arguments are certainly valid for the case $\varepsilon=\mbox{const.}$ For viscoelastic material properties, however, with decaying thermal velocity the collisions occur less and less dissipatively since $\varepsilon_v$ approaches one with propagating time. Hence the question arises whether cluster formation in Granular Gases of viscoelastic particles occurs as a transient phenomenon only. This question is difficult and there is no conclusive answer yet, although the linear stability analysis \cite{BrilliantovPoeschel:2001roy} seems to support the hypothesis that cluster formation is only a transient phenomenon in Granular Gases of viscoelastic particles.

The numerical simulation of clustering Granular Gases is problematic since clusters grow rapidly and at a certain time their extension is comparable with the system size. From this instant on the simulation becomes invalid since the cluster interacts with itself via the periodic boundary conditions. For the case of viscoelastic particle properties $\varepsilon=\varepsilon(g)$ we expect that clusters dissolve at late times when the typical coefficient of restitution $\varepsilon\left(v_T\right)$ approaches one. To prove this by MD simulations we need a system which is large enough to assure that at no time the cluster size reaches the system dimensions. At the moment we are able to simulate systems up to about $N=3\times 10^6$ particles, which seems to be far below the necessary size.

\begin{figure}[htbp]
\centerline{\includegraphics[width=1\textwidth]{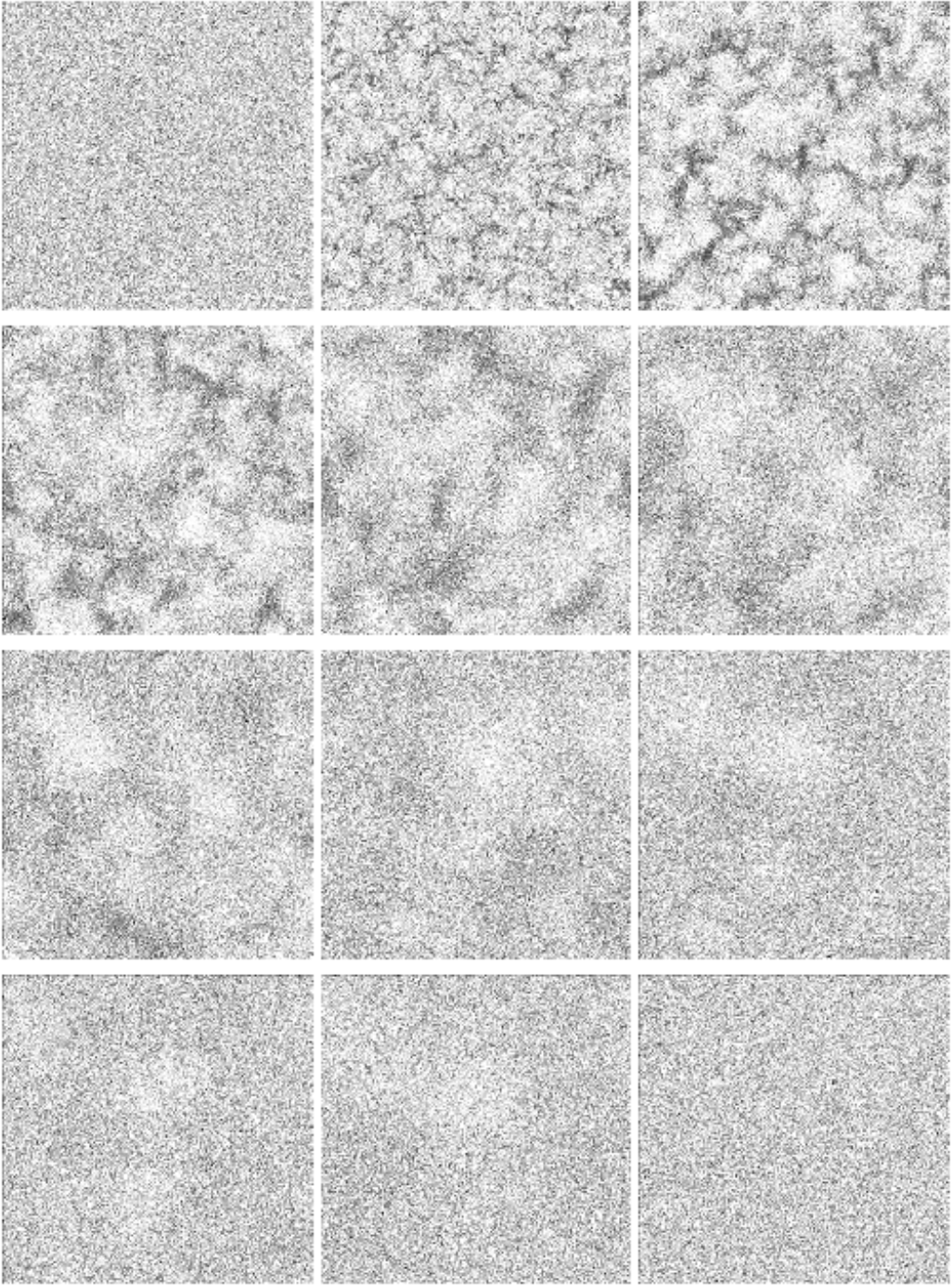}}
%\ifthenelse{\equal{\pdf}{y}}
%{  \centerline{\includegraphics[width=0.33\textwidth]{figs/0.jpg}~\includegraphics[width=0.33\textwidth]{figs/1.jpg}~\includegraphics[width=0.33\textwidth]{figs/2.jpg}}\vspace*{0.2cm}
%  \centerline{\includegraphics[width=0.33\textwidth]{figs/3.jpg}~\includegraphics[width=0.33\textwidth]{figs/4.jpg}~\includegraphics[width=0.33\textwidth]{figs/5.jpg}}\vspace*{0.2cm}
%  \centerline{\includegraphics[width=0.33\textwidth]{figs/6.jpg}~\includegraphics[width=0.33\textwidth]{figs/10.jpg}~\includegraphics[width=0.33\textwidth]{figs/15.jpg}}\vspace*{0.2cm}
%  \centerline{\includegraphics[width=0.33\textwidth]{figs/20.jpg}~\includegraphics[width=0.33\textwidth]{figs/25.jpg}~\includegraphics[width=0.33\textwidth]{figs/long-20.jpg}}\vspace*{0.2cm}}
%{  \centerline{\includegraphics[width=0.33\textwidth]{figs/0.eps}~\includegraphics[width=0.33\textwidth]{figs/1.eps}~\includegraphics[width=0.33\textwidth]{figs/2.eps}}\vspace*{0.2cm}
%  \centerline{\includegraphics[width=0.33\textwidth]{figs/3.eps}~\includegraphics[width=0.33\textwidth]{figs/4.eps}~\includegraphics[width=0.33\textwidth]{figs/5.eps}}\vspace*{0.2cm}
%  \centerline{\includegraphics[width=0.33\textwidth]{figs/6.eps}~\includegraphics[width=0.33\textwidth]{figs/10.eps}~\includegraphics[width=0.33\textwidth]{figs/15.eps}}\vspace*{0.2cm}
%  \centerline{\includegraphics[width=0.33\textwidth]{figs/20.eps}~\includegraphics[width=0.33\textwidth]{figs/25.eps}~\includegraphics[width=0.33\textwidth]{figs/long-20.eps}}\vspace*{0.2cm}}
  \caption{Snapshots of a simulation of $N=10^5$ particles ($\varepsilon^*=0.6$, $g^*=0.1$, $\eta=0.1$). The pictures  show the system after 0, 50, 100 (first row), 150, 200, 250 (second row), 300, 500, 750 (third row), 1000, 1250 and 5000 (last row) collisions per particle. The initial temperature is $T_0=1$ which corresponds to the thermal velocity $v_T=\sqrt{2}$. Starting with homogeneously distributed particles we observe cluster formation up to a certain cluster size. At late times the system returns to the homogeneous cooling state.}
  \label{fig:snaps}
\end{figure}

We have simulated a system of $n=10^5$ particles which collide according to Eq. (\ref{eq:epsstep}). This collision rule reveals the same asymptotic behavior as the rule (\ref{eq:epsvisco}) for the collision of viscoelastic spheres, $\lim_{g\to 0} \varepsilon = 1$, nevertheless, the law is drastically simplified. Snapshots of the simulation are shown in Fig. \ref{fig:snaps}. 

Starting with a uniform distribution after a short period of homogeneous cooling density inhomogeneities appear which grow up to a certain typical size. The collision frequency in dense regions is larger than in dilute regions which leads to decreasing thermal velocity. In the due of time more and more collisions inside the clusters occur with relative velocities $g<g^*$, i.e., elastically. Therefore, the clusters dissolve. At late times the system returns to the homogeneous cooling state. 

Why do we not observe system spanning clusters as in the case of $\varepsilon=\mbox{const.}$? The reason is the dynamics of cluster growth. To grow a cluster of a given size the system needs a certain time which is determined by the density and by the value of the coefficient of restitution. We start our simulation at a temperature which corresponds to a thermal velocity which is well above $g^*$, i.e., the gas has essentially the properties of a Granular Gas with $\varepsilon=\mbox{const.}$ Therefore, clusters begin to form and grow due to the described pressure instability. At a certain time the typical velocity inside the clusters falls below $g^*$, i.e., from this moment on the clusters start to dissolve. This time corresponds to a typical cluster size which is significantly smaller than the system size. Therefore, we do not observe clusters beyond a certain typical size.

According to a similar argumentation we expect clusters in Granular Gases of viscoelastic particles to dissolve too. Due to numerical limitations at present time it is not possible to perform the corresponding simulations without encountering the problem of boundary conditions. 

In the recent study \cite{NieBenNaimChen2002}, where the restitution coefficient of the form (\ref{eq:epsstep}) was used to avoid the inelastic collapse, the dissolution of clusters has not been detected. The reason for the persistent development of clusters in \cite{NieBenNaimChen2002} was a very small value of $g^*$, which was at least 100 times smaller than the thermal velocity of particles during the simulations. In our numerical experiments the clusters started to dissolve when the thermal velocity
became comparable with $g^*$. This regime has not been reached in simulations of \cite{NieBenNaimChen2002}.

\section{Conclusion}
We have investigated a model Granular Gas of particles which collide with a stepwise coefficient of restitution which mimics the coefficient of restitution of viscoelastic particles: The particles collide elastically ($\varepsilon=1$) for small impact velocities and with $\varepsilon=\mbox{const.}<1$ if the relative particle velocity exceeds a certain value ($g>g^*$). This simplified dependence of the restitution coefficient reflects the main feature of viscoelastic particles that collisions tend to occur elastically for decreasing impact velocity. We analyze the temperature decay and the dynamics of cluster formation. We observe that clustering occurs in this model only as a transient process. The important property of the collision model is that the clusters dissolve before system spanning clusters appear. Hence, this model allows to investigate the total process without the drastic influence of the boundary conditions. The results of our study supports the hypothesis that clustering occurs as a transient process for realistic gases of viscoelastic particles, as was predicted before by means of the linear stability analysis. 

%\bibliography{EPSSTEP}

\begin{thebibliography}{10}
\expandafter\ifx\csname url\endcsname\relax
  \def\url#1{\texttt{#1}}\fi
\expandafter\ifx\csname urlprefix\endcsname\relax\def\urlprefix{URL }\fi

\bibitem{GoldshteinShapiro1:1995}
A.~Goldshtein, M.~Shapiro, Mechanics of collisional motion of granular
  materials. {P}art 1: {G}eneral hydrodynamic equations, J. Fluid Mech. 282
  (1995) 75.

\bibitem{NoijeErnst:1998}
T.~P.~C. van Noije, M.~H. Ernst, Velocity distributions in homogeneous granular
  fluids: the free and the heated case, Granular Matter 1 (1998) 57.

\bibitem{BrilliantovPoeschelStability:2000}
N.~V. Brilliantov, T.~P\"oschel, Deviation from {M}axwell distribution in
  granular gases with constant restitution coefficient, Phys. Rev. E 61 (2000)
  2809.

\bibitem{EsipovPoeschel:1995}
S.~E. Esipov, T.~P\"oschel, The granular phase diagram, J. Stat. Phys. 86
  (1997) 1385.

\bibitem{BrilliantovPoeschel:1998d}
N.~V. Brilliantov, T.~P\"oschel, Self-diffusion in granular gases, Phys. Rev. E
  61 (2000) 1716.

\bibitem{BreyRuizMonteroCuberoGarcia:2000}
J.~J. Brey, M.~J. Ruiz-Montero, D.~Cubero, R.~Garcia-Rojo, Self-diffusion in
  freely evolving granular gases, Physics of Fluids 12 (2000) 876.

\bibitem{PoeschelLuding:2001}
T.~P\"oschel, S.~Luding (Eds.), Granular Gases, Vol. 564 of Lecture Notes in
  Physics, Springer, Berlin, 2001.

\bibitem{GoldhirschZanetti:1993}
I.~Goldhirsch, G.~Zanetti, Clustering instability in dissipative gases, Phys.
  Rev. Lett. 70 (1993) 1619.

\bibitem{BritoErnst:1998}
R.~Brito, M.~H. Ernst, Extension of {H}aff's cooling law in granular flows,
  Europhys. Lett. 43 (1998) 497.

\bibitem{BridgesHatzesLin:1984}
F.~G. Bridges, A.~Hatzes, D.~N.~C. Lin, Structure, stability and evolution of
  {S}aturn's rings, Nature 309 (1984) 333.

\bibitem{Tanaka}
T.~Tanaka, T.~Ishida, Y.~Tsuji, Direct numerical simulatin of granular plug
  flow in a horizontal pipe: {T}he case of cohesionless particles (in
  {J}apanese, for an english presentation of this work
  see~\cite{Taguchi:1992JDP}), Trans. Jap. Soc. Mech. Eng. 57 (1991) 456.

\bibitem{Ramirez:1999}
R.~Ram\'{\i}rez, T.~P{\"o}schel, N.~V. Brilliantov, T.~Schwager, Coefficient of
  restitution for colliding viscoelastic spheres, Phys. Rev. E 60 (1999) 4465.

\bibitem{Hertz:1882}
H.~Hertz, {\"U}ber die {B}er\"uhrung fester elastischer {K}\"orper, J. f. reine
  u. angewandte Math. 92 (1882) 156.

\bibitem{BrilliantovSpahnHertzschPoeschel:1994}
N.~V. Brilliantov, F.~Spahn, J.-M. Hertzsch, T.~P\"oschel, A model for
  collisions in granular gases, Phys. Rev. E 53 (1996) 5382.

\bibitem{MorgadoOppenheim:1997}
W.~A.~M. Morgado, I.~Oppenheim, Energy dissipation for quasielastic granular
  particle collisions, Phys. Rev. E 55 (1997) 1940.

\bibitem{KuwabaraKono:1987}
G.~Kuwabara, K.~Kono, Restitution coefficient in a collision between two
  spheres, Jpn. J. Appl. Phys. 26 (1987) 1230.

\bibitem{SchwagerPoeschel:1998}
T.~Schwager, T.~P\"oschel, Coefficient of restitution of viscous particles and
  cooling rate of granular gases, Phys. Rev. E 57 (1998) 650.

\bibitem{BrilliantovPoeschel:2001roy}
N.~V. Brilliantov, T.~P\"oschel, Hydrodynamics of granular gases of
  viscoelastic particles, Phil. Trans. R. Soc. Lond. A 360 (2001) 415.

\bibitem{NieBenNaimChen2002}
X.~Nie, E.~Ben-Naim, S.~Y. Chen, Dynamics of freely cooling granular gases,
  Phys. Rev. Lett. 89 (2002) 204301.

\bibitem{Haff:83}
P.~K. Haff, Grain flow as a fluid-mechanical phenomenon, J. Fluid Mech. 134
  (1983) 401.

\bibitem{BrilliantovPoeschel:2000visc}
N.~V. Brilliantov, T.~P\"oschel, Velocity distribution of granular gases of
  viscoelastic particles, Phys. Rev. E 61 (2000) 5573.

\bibitem{Corless:1996}
R.~M. Corless, G.~H. Gonnet, D.~E.~G. Hare, D.~J. Jeffrey, D.~E. Knuth, On the
  {L}ambert {W} {F}unction, Adv. Comput. Math. 5 (1996) 329.

\bibitem{Taguchi:1992JDP}
Y.~Taguchi, Powder turbulence: {D}irect onset of turbulent flow, J. Physique 2
  (1992) 2103.

\end{thebibliography}

\end{document}